\documentclass[12pt,reqno]{article}
\usepackage{amsmath,hyperref,amssymb}
\usepackage{graphicx,url}

\allowdisplaybreaks \numberwithin{equation}{section}
\renewcommand{\Large}{\large} 

\DeclareSymbolFont{AMSa}{U}{msa}{m}{n}
\DeclareSymbolFont{AMSb}{U}{msb}{m}{n}
\DeclareMathSymbol{\fieldR}{\mathalpha}{AMSb}{"52}

\begin{document}

\begin{flushright} \small
ITP--UU--12/24 \\ SPIN--12/22
\end{flushright}
\bigskip
\begin{center}

{\Large \bfseries Static nonextremal AdS$_4$ black hole solutions}\\[8mm]
Chiara Toldo and  Stefan Vandoren \\[8mm]
{\small\slshape
Institute for Theoretical Physics \emph{and} Spinoza Institute, \\
Utrecht University, 3508 TD Utrecht, The Netherlands \\

\medskip

\medskip

{\upshape\ttfamily  C.Toldo@uu.nl, S.J.G.Vandoren@uu.nl}\\[3mm]}
\end{center}
\vspace{8mm}
\hrule

\bigskip 
\centerline{\bfseries Abstract}
\begin{changemargin}{1.2cm}{1.2cm} 
We find new static nonextremal black hole solutions that asymptote to AdS$_4$ in  $D=4$  gauged $\mathcal{N}=2$ supergravity. Solutions include  electric and magnetic black holes with constant scalars that in the BPS limit reduce to naked singularities, but also magnetic black holes with running scalars that at extremality reduce to BPS black holes with finite horizon area. For all these solutions we compute area product formulae and show they are independent of the mass. Finally, we also find new examples of nonextremal magnetic black branes.
 \end{changemargin}

\hrule

\bigskip

\tableofcontents

\section{Introduction}

Black holes are between the most interesting objects for probing a theory of quantum gravity. The correspondence between the laws of black hole mechanics and thermodynamics raises a lot of challenging questions. Especially for black holes in anti-de Sitter (AdS) spacetimes, this correspondence is not well understood. In particular, a microscopic explanation of the entropy area law of four dimensional AdS$_4$ BPS black holes is still an open problem. But already at the classical level, which is the focus of our analysis here, there are still a number of unresolved issues, such as the nature of the attractor mechanism for BPS or extremal solutions, and the role played by the scalar fields in gauged supergravity. Moreover, sofar, all known nonextremal AdS$_4$ solutions in gauged supergravity develop naked singularities in the BPS limit and are therefore ultimately unstable. One of our aims is to present an example of a class of static AdS$_4$ black holes whose  BPS limit is smooth and stable.

In this paper we consider four-dimensional $\mathcal{N}=2$ gauged supergravity models that allow for asymptotically AdS$_4$ black holes. These provide gravitational backgrounds that can be relevant for the AdS$_4$/CFT$_3$ correspondence \cite{Aharony:2008ug}. A classification of AdS$_4$ black holes in gauged supergravities, in the supersymmetric  \cite{Cacciatori:2008ek,Klemm:2010mc,Hristov:2010eu,Meessen:2012sr} and in the nonextremal cases, is still in progress. So far some examples of supersymmetric AdS$_4$ black holes were found: rotating with electric \cite{kostelecky} or magnetic charges \cite{Klemm:2011xw}, or static with spherical symmetry magnetic or dyonic \cite{klemm-adsBH,Dall'Agata:2010gj,kiril_stefan}\footnote{For other supersymmetric solutions with different topologies of event horizon see \cite{ortin,Caldarelli:1998hg}.}. In particular, in this last case the presence of magnetic charge and running scalars allow the existence of static extremal BPS black holes, while instead static BPS solutions with only electric charges seem to always generate naked singularities \cite{sabra}. 

For what concerns nonextremal AdS$_{4}$ solutions of gauged $D=4$ supergravity, some examples were found for instance in \cite{cvetic_rotating,Chow:2010sf,Chow:2010fw}; they are rotating black holes.
Static nonextremal configurations were instead  found in \cite{duff_liu} by means of the same technique as in \cite{Duff:1996hp}\footnote{The technique was previously and successfully applied to $D=5$ AdS black holes by \cite{cvetic5d}.}. A common feature of these static $D=4$ black holes is that in the  BPS limit (when it exists) they reduce to naked singularities. 

The main aim of this paper is to provide further examples of static nonextremal black holes as deformations of BPS solutions. Examples contain electric and magnetic black holes with constant scalars, that reduce in the BPS limit to the solutions found in \cite{sabra,chamseddine}, that are singular.
In the case of running scalars, we retrieve the solutions of \cite{duff_liu} as a subset in the electric case and furthermore we find a nonextremal generalization of the magnetic static solution found in \cite{kiril_stefan}. This last solution is new and of particular interest since it can be used to study further properties of mAdS$_4$  \cite{ours}, the ground state underlying the magnetic BPS black hole solution. Furthermore, the BPS limit is a regular black hole with  nonzero area of the event horizon, contrary to what happens in the electric case.

We work in gauged $\mathcal{N}=2$ supergravity in presence of abelian gaugings with Fayet-Iliopoulos terms without hypermultiplets. This is the simplest model in which scalar fields are present. These scalars are neutral under the gauge group; the only charged particles are the two gravitinos of opposite electric charge.
The explicit black hole solutions we present are given for the particular case of prepotential $F=-2i\sqrt{X^0 (X^1)^3}$, which is one for which a string/M-theory embedding exists, but we believe that qualitative features should appear also in presence of other prepotentials.

As the last example, we apply our technique to the magnetic BPS black branes discussed in \cite{brane}, generating a new nonextremal magnetic black brane. These magnetic brane solutions might have interesting dual field theories on the boundary.

For all the solutions we present, the mass and the central charge (where present) is computed given  the formulas for nonminimal supergravity found in \cite{kiril}. We will comment later on the outcome of these formulas and on BPS bounds.
Furthermore, for all black hole solutions it is verified explicitely that the product of the areas of the horizons depends just on the charges and not on (parameters depending on) the mass.

{\bf Note added}: The results of this paper were presented during the Carg\`ese Summer School on String Theory, June 4-16, 2012. After that, during the write-up of our work, the paper arXiv:1207.2678 by D. Klemm and O. Vaughan appeared \cite{K-V}. There is some overlap with our analysis on the construction of the non-extremal magnetic solutions for the $t^3$ model. In \cite{K-V}, also other models were studied, whereas here, we focussed on the computation of the masses and central charges, and the product area formulae in the $t^3$ model. It would be interesting to repeat this analysis for the additional models considered in \cite{K-V}.

\section{Preliminaries}

Extensive details about abelian $\mathcal{N}=2$ gauged supergravity can be found in \cite{abcd}, whose notations are mostly adopted here. In the context of black hole physics, these models were also discussed in \cite{klemm-adsBH,Dall'Agata:2010gj,kiril_stefan}.

As the gauge group is abelian, the  only charged fields are the gravitinos, meanwhile the
vector multiplet scalars are neutral (this is usually called Fayet-Iliopoulos (FI) gauging). The gauge fields
couple to the gravitini through a linear combination of the graviphoton and the $n_V$
vectors from the vector multiplets, $\xi_{\Lambda}A_{\mu}^{\Lambda}$, with $\Lambda=0,..., n_V$. The constants $\xi_{\Lambda}$ are called FI parameters. The electric charges of the gravitinos are then denoted by
\begin{equation}\label{grav-charge}
e_\Lambda \equiv g \xi_\Lambda\ .
\end{equation}
The two gravitinos have opposite charges $\pm e_\Lambda$ under the gauge group $U(1)^{n_V+1}$. Since the FI parameters determine the electric charge, we assume them to be quantized.

\subsection{Supergravity lagrangian and equations of motion}
In the conventions of \cite{kiril_stefan}, the bosonic part of the action for abelian gauged $D=4$ $\mathcal{N}=2$ supergravity with $n_V$ vector multiplets and in absence of hypermultiplets is:
\begin{equation}
S=\int {\rm d}^4x\, e \left[ \frac12 R +g_{i \bar{\jmath}} \partial^{\mu}z^{i} \partial_{\mu} \bar{z}^{\bar{\jmath}} + I_{\Lambda \Sigma} F_{\mu \nu}^{\Lambda} F^{\mu \nu |\Sigma} +\frac12 R_{\Lambda \Sigma} \epsilon^{\mu \nu \rho \sigma} F_{\mu \nu}^{\Lambda} F_{\rho \sigma}^{\Sigma} - g^2 V(z, \bar{z})\right] \, ,
\end{equation}
with $\Lambda , \Sigma= 0,1... , n_V$, $i,j=1,..., n_V$. The imaginary and the real part of the period matrix $\mathcal{N}_{\Lambda \Sigma}$ (more details about special K\"ahler quantities are given in the Appendix) are respectively denoted by $I_{\Lambda \Sigma}$ and $R_{\Lambda \Sigma}$. The complex scalars $z^i$ are written in terms of the holomorphic symplectic sections $(X^{\Lambda},F_{\Lambda})$. Furthermore, the scalar potential has this form:
\begin{equation}
V= (g^{i \bar{\jmath}} f_i^{\Lambda}\bar{f}_{\bar{\jmath}}^{\Sigma}-3\bar{L}^{\Lambda}L^{\Sigma}) \xi_{\Lambda} \xi_{\Sigma} \,.
\end{equation}

Einstein's equations then are:
$$
-(R_{\mu \nu} - \frac12 g_{\mu \nu} R) = g_{\mu \nu} g^2 V(z, \bar{z}) - g_{\mu \nu} \partial^{\sigma}z^{i} \partial_{\sigma} \bar{z}^{\bar{\jmath}} g_{i \bar{\jmath}}+2  g_{i \bar{\jmath}} \partial_{\mu}z^{i} \partial_{\nu}\bar{z}^{\bar{\jmath}} +
$$
\begin{equation}
- I_{\Lambda \Sigma} g_{\mu \nu} F_{\rho \sigma}^{\Lambda} F^{\rho \sigma | \Sigma} +4 I_{\Lambda \Sigma}  F_{\mu \alpha}^{\Lambda} {F_{\nu}}^{\alpha|\Sigma}\,.
\end{equation}

The equations of motion for the scalar fields $z^i$ read:
\begin{equation}
- g_{i \bar{\jmath}} \partial_{\mu}(e \partial^{\mu}\bar{z}^{\bar{\jmath}})- e\frac{\partial g_{i \bar{k}}}{\partial \bar{z}^{\bar{\jmath}}} \partial^{\mu}\bar{z}^{\bar{\jmath}} \partial_{\mu} \bar{z}^{\bar{k}}+ e \frac{\partial I_{\Lambda \Sigma}}{\partial z^{i}} F_{\rho \sigma}^{\Lambda} F^{ \rho \sigma |\Sigma} + \frac{e}{2} \frac{\partial R_{\Lambda \Sigma}}{\partial z^{i}} \epsilon^{\mu \nu \rho \sigma} F_{\mu \nu}^{\Lambda} F_{\rho \sigma}^{\Sigma} - e g^2 \frac{\partial V}{\partial z^{i}}=0\,,
\end{equation}
and the Maxwell's equations for the vector fields $A_{\nu}^{\Lambda}$ are:
\begin{equation}
\partial_{\mu}(e F^{\mu \nu \,| \Sigma} I_{\Sigma \Lambda} - \frac{e}{2} \epsilon^{\mu\nu\rho\sigma}F_{\rho \sigma}^{\Sigma} R_{\Sigma \Lambda})=0\,.
\end{equation}

\subsection{Plugging in the ansatz}
At this point we  restrict ourself to the static and spherically symmetric ansatz 
\begin{equation}\label{ansatz_metric}
{\rm d}s^2= U^2(r) {\rm d}t^2 - \frac{1}{U^2(r)} {\rm d}r^2- h^2(r) ( {\rm d}\theta^2+ \sin^2 \theta {\rm d}\varphi^2)\,.
\end{equation}
Due to spherical symmetry, the Bianchi identity and the Maxwell equation, the form of  the field strengths is restricted to be 
\begin{equation}\label{ansatz_fs}
F_{tr}^{\Lambda}=- \frac{1}{2h^2(r)} I^{\Lambda \Sigma} (R_{\Sigma \Gamma}p^{\Gamma}- q_{\Sigma}) \,, \qquad
F_{\theta \varphi}^{\Lambda}= \frac12 p^{\Lambda} \sin \theta \,,
\end{equation}
while all other components vanish. 
Here $I^{\Lambda \Sigma}$ is the inverse of $I_{\Lambda \Sigma}$, and the magnetic and electric charges satisfy
\begin{equation}
p^{\Lambda}=- \frac{1}{4\pi}\int_{S^2_{\infty}} F^{\Lambda}\,, \qquad q_{\Lambda}= -\frac{1}{4\pi} \int_{S^2_{\infty}} G_{\Lambda}\,,
\end{equation}
with ($\mathcal{L}$ is the Lagrangian density, i.e.  $ S= \int {\rm d}^4x \, e \, \mathcal{L}$)
\begin{equation}
G_{\mu \nu | \Lambda}= \epsilon_{\mu \nu \rho \sigma}\frac{\partial \mathcal{L}}{\partial F_{\rho \sigma}^{\Lambda}}\,.
\end{equation}
The matrices $I_{\Lambda \Sigma}$ and $R_{\Lambda \Sigma}$ depend on the specific model taken into consideration and for the moment we keep $q_{\Lambda}$ and $p^{\Lambda}$ unconstrained.

The scalar field equation reduces to:
$$
\frac{1}{h^2(r)} g_{i \bar{\jmath}} \partial_r \left( h^2(r)  U^2(r) \partial_r \bar{z}^{\bar{\jmath}} \right) - \frac{ \partial{g_{i \bar{\jmath}}}}{\partial \bar{z}^{\bar{k}}} \partial_r \bar{z}^{\bar{\jmath}} \partial^r \bar{z}^{\bar{k}} + \frac{\partial I_{\Lambda \Sigma}}{\partial z^i} F_{\mu \nu}^{\Lambda}F^{\mu \nu | \Sigma}+
$$
\begin{equation}\label{eomscalars}
 + \frac12 \frac{\partial R_{\Lambda \Sigma}}{\partial z^{i}} \epsilon^{\mu \nu \rho \sigma} F_{\mu \nu}^{\Lambda} F_{\rho \sigma}^{\Sigma} -g^2 \frac{\partial V}{\partial z^i}=0\,.
\end{equation}

The relevant components of the Einstein's equation yield:\\
$tt$ component:
\begin{equation}
- \frac{\left( -1+2 h \, \, h' \, U' U+ U^2 (h'^2+ 2h\, h'') \right)  }{h^2}= g^2 V(z, \bar{z}) -\partial^r z^i \partial_r \bar{z}^{\bar{\jmath}}g_{i \bar{\jmath}} - \frac{V_{BH}}{h^4}\,,
\end{equation}
$rr$ component:
\begin{equation}
-\frac{-1+U^2 h'^2+2 h\,h' \, U' U}{h^2} = g^2 V(z, \bar{z}) + \partial^r z^i \partial_r \bar{z}^{\bar{\jmath}}g_{i \bar{\jmath}} - \frac{V_{BH}}{h^4}\,,
\end{equation}
$\theta \theta$ component:
\begin{equation}
-\frac{ \left( h U'^2 +U^2 h'' +U(2 h' U' +h U'')   \right)}{h} = g^2 V(z, \bar{z})- \partial^r z^i \partial_r \bar{z}^{\bar{\jmath}}g_{i \bar{\jmath}} + \frac{V_{BH}}{h^4}\,,
\end{equation}
where
\begin{equation}\label{VBH}
V_{BH}\, (z,\bar{z},q_{\Lambda},p^{\Lambda})= -\frac12 \left(\begin{array}{cc}
p^{\Lambda} & q_{\Lambda}
\end{array}
\right)  \left(
\begin{array}{cc}
  I_{\Lambda \Sigma}+ R_{\Lambda \Gamma} I^{\Gamma \Delta} R_{\Delta \Sigma} & - R_{\Lambda \Gamma} I^{ \Gamma \Sigma} \\
- I^{\Lambda \Gamma }R_{\Gamma \Sigma}  & I^{\Lambda \Sigma} \\
\end{array}
\right)
\left(
\begin{array}{c}
p^{\Sigma} \\
q_{\Sigma}
\end{array}
\right)\,.
\end{equation}

The $\varphi \varphi$ component gives the $\theta \theta$ one multiplied by $\sin^2\theta$; all other nondiagonal components are trivial.

We now manipulate the equations in order to get simpler ones, along the same lines as \cite{ferrara}.
Adding $tt$ and $rr$ we get
\begin{equation}\label{einst1}
-2 \frac{h''}{h}= 2g_{i \overline{\jmath}} \partial_r z^{i} \partial_r \bar{z}^{\bar{\jmath}}\,,
\end{equation}
while adding $rr$ and $\theta \theta$ we obtain
\begin{equation}\label{einst2}
\frac{1- (U^2\,h^2)''/2}{h^2}  = 2 g^2 V(z,\overline{z})\,.
\end{equation}
Finally, $\theta \theta - tt$ gives
\begin{equation}\label{einst3}
-1+ U^2 h'^2 +h h'' U^2 - h^2 U'^2 -h^2 U U'' = \frac{2 V_{BH}}{h^2}\,.
\end{equation}
Furthermore one can show that for nonconstant scalars if the Einstein's
$rr,\,\, tt$ component and the equation of motion of the scalars are satisfied, the $\theta \theta$ component is satisfied too. In the explicit examples we restrict to the case of a single scalar. In the case of one nonconstant scalar we solve first the three Einstein's equations, then we verify that the scalar equation does not provide any further constraints. For constant scalar configurations, however, this is not true and one needs to solve the scalar equation. We treat that case separately in the next section.

\section{Constant scalars solutions}

We now  want to construct nonextremal solutions with constant scalars, in presence of an  arbitrary prepotential. For the moment, we keep constant complex scalar fields, but will later restrict to real ones.

First of all we analyze the equation \eqref{einst1}. The right hand side is identically zero, so we integrate twice and find $h(r)$:
\begin{equation}\label{const_b}
h(r)= a\,r +b\,, \qquad a , b \,{\rm \,\,\, constant}. 
\end{equation}
We keep for the moment also the  integration constant $b$, in order to deal with the solution in its full generality\footnote{Solutions of the form \eqref{const_b} are found for instance when we impose constant scalars and sections $X^{\Lambda}$ proportional to each other. When we impose also constant sections, we find solutions with $b=0$.}.

Analyzing the scalar equation of motion \eqref{eomscalars}, then, due to the different radial dependence, the two remaining terms have to vanish separately
\begin{equation}\label{constr_quadratic}
 \frac{\partial I_{\Lambda \Sigma}}{\partial z^i} F_{\mu \nu}^{\Lambda}F^{\mu \nu | \Sigma} + \frac12 \frac{\partial R_{\Lambda \Sigma}}{\partial z^{i}} \epsilon^{\mu \nu \rho \sigma} F_{\mu \nu}^{\Lambda} F_{\rho \sigma}^{\Sigma} =0\,, 
\end{equation}
\begin{equation}\label{deV}
\frac{\partial V}{\partial z^i}=0\,,
\end{equation}
and complex conjugates. This sets the scalars at their constant value that extremizes the scalar potential
\begin{equation}
z^i=z_*^i \qquad {\rm such \,\,\, that} \qquad \frac{\partial V}{\partial z^i}|_{z^i=z_*^i } =0
\end{equation}

and imposes the constraint \eqref{constr_quadratic}, that is quadratic in the electric and magnetic charges.

We now integrate the equation \eqref{einst2}, that gives the form of the warp factor
\begin{equation}\label{formU}
U^2(r)= \frac{-g^2 V_* \left(\frac{a^2 r^4}{3} +\frac{4 a\,b \,r^3}{3} +2 b^2 r^2 \right) +r^2- \mu r +Q }{(ar+b)^2}\,.
\end{equation}
When $V_*=V(z_*^{i}, \bar{z}_*^{\bar{\imath}})$ is negative the solution asymptotes to AdS$_4$. The parameters $\mu$, $Q$ are further (real) integration constants. 
Finally  equation \eqref{einst3} gives another constraint:
\begin{equation}\label{constr_2}
b^4 g^2 V_* -b^2 -a\,b\, \mu -a^2Q=V_{BH}\,,
\end{equation}
where $V_{BH}$ is given in \eqref{VBH}, and is here evaluated at the critical point $z_*$. The most general solutions with constant scalars is then of the form \eqref{const_b} \eqref{formU} and has to satisfy the constraints \eqref{constr_quadratic} \eqref{constr_2}.

We will now specialize to electric and magnetic black holes with constant scalars and prepotential $F=-2i \sqrt{X^0 (X^1)^3}$. The general solution is characterized by two magnetic chages $p^0$, $p^1$ and two electric ones $q_0$, $q_1$, and one scalar $z=\frac{X^1}{X^0}$ that we assume to be  real for simplicity of the calculations presented below. The special K\"ahler quantities for this prepotential are given in the Appendix. Furthermore, the positivity of the K\"ahler metric requires the scalar field to be positive.

In the explicit examples, and for similarity with the Reissner-Nordstr\"om case, we also assume $b=0$. Equation
 \eqref{deV} is then satisfied if we impose the constant value of the scalar field
\begin{equation}
z= z_{*} = \frac{3 \xi_0}{\xi_1}\,,
\end{equation}
that gives the value of the potential 
\begin{equation}\label{vstar}
V_{*}= - \frac{2}{\sqrt3} \sqrt{\xi_0 \xi_1^3}\ .
\end{equation}  
This implies that the cosmological constant is negative, 
\begin{equation}\label{cc}
\Lambda = 3 V_* g^2\ ,
\end{equation}
such that we have an AdS$_4$ extremum (in particular it is a local maximum). For the field $z$ to have positive kinetic terms, $z$ should be positive and the FI terms then must have the same sign. We choose positive FI terms without loss of generality, since the lagrangian remains invariant under the transformation for which both $\xi_0$ and $\xi_1$ change sign, provided the gravitinos flip charge.

In what follows we will analyze separately purely electric and purely magnetic black holes. We require the new nonextremal solutions to be deformations of the BPS states found respectively in \cite{sabra} and \cite{chamseddine}; this imposes a constraint on the charges present in the configurations.
 For these nonextremal black holes we compute the mass and the other charges, given the formulas of \cite{kiril}. The Hawking temperature can be computed along the same lines as \cite{romans}.

One can furthermore compare the BPS bound with the extremality bound for these new solutions.
First we rescale the radial and time coordinates, $r\rightarrow r/a, t\rightarrow at$ in \eqref{formU} with $b=0$ such that we retrieve the familiar form
\begin{equation}
U^2(r)=1-\frac{2m}{r} +\frac{Z^2}{r^2}+g'^2r^2\,.
\end{equation}
The parameters $m$, $Z$, $g'$ are related to the ones of \eqref{formU} through
\begin{equation}
m=\frac12 a \mu \,, \qquad Z^2=a^2 Q\,, \qquad g'^2=\frac{|V_{*}|g^2}{3a^2}\,.
\end{equation}
To compute the value of $\mu$ appearing in \eqref{formU} at extremality, $\mu_{extr}(Q,g,V_{*})$, we require that the function $U(r)$ and also its derivative with respect to $r$ vanish at the horizon:
\begin{equation}\label{cond_extr}
U=0 \,\,\,\,\,\,\, {\rm and} \,\,\,\,\,\,\, \frac{dU}{dr}=0 \,\,\,\,\,\,\,\,{\rm at\,\, the\,\, horizon}\,.
\end{equation}
Adding these two equations we get a condition on the radius of the event horizon:
\begin{equation}
r_{hor}^2= \frac{-1+\sqrt{1+4g^2|V_{*}| Q}}{2g^2 |V_{*}|} a^2\,,
\end{equation}
and plugging in back in the first of eqq. \eqref{cond_extr} we finally get
\begin{equation}\label{extr}
\mu_{extr}= \frac{\sqrt2}{3 \, g \sqrt{|V_{*}|}}(\sqrt{1+4g^2|V_{*}|  Q}+2)(\sqrt{1+4g^2|V_{*}|  Q}-1)^{1/2}\,.
\end{equation}
The same expression (in terms of $g$, $Z$) was also obtained in \cite{Caldarelli:1998hg}. For $\mu \geq \mu_{extr}$ the solution represents a black hole, otherwise a naked singularity.
We remind that for the solutions obtained in \cite{romans} in the realm of minimal gauged supergravity the BPS bound lies below the extremality bound; this means that supersymmetric solutions are naked singularities. In the examples with constant scalars that follow below, we will see that the situation remains the same.

\subsection{Explicit examples}
\subsubsection{Electric solution}
For this configuration the magnetic charges are set to zero, $p^0=p^1=0$, meanwhile the two electric charges are unconstrained. We remind that the real scalar is fixed at the value $z=\frac{3\xi_0}{\xi_1}$, so that \eqref{deV} is satisfied. We construct a solution of the form \eqref{formU} with $b=0$ and $a=\frac{\sqrt{2} (\xi_0 \xi_1^3)^{1/4}}{3^{3/4}}$. This cumbersome factor is  somewhat  convenient for the comparison of these solutions with the ones with running scalars of Section 3. The metric components then become
\begin{equation}
h(r)=\frac{\sqrt{2} (\xi_0 \xi_1^3)^{1/4}r}{3^{3/4}}\,, \qquad
U^2(r)= \frac{ 3 \sqrt3  \left(\frac{4}{27} \xi_0 \xi_1^3 g^2 r^2 +1 - \frac{\mu}{r} + \frac{Q}{r^2}  \right)}{ 2 \sqrt{\xi_0 \xi_1^3}}\,.
\end{equation}
Equation \eqref{constr_2} then imposes a constraint on $Q$
\begin{equation}\label{Qel}
Q=\frac14 \left( \frac{q_0^2}{\xi_0^2}+ 3 \frac{q_1^2}{\xi_1^2}\right)\,,
\end{equation}
while \eqref{constr_quadratic} gives
\begin{equation}\label{constrscalars}
q_0 = \pm \xi_0 \frac{q_1}{\xi_1}\,,\qquad \rightarrow \qquad Q=\frac{q_1^2}{\xi_1^2}\ .
\end{equation}
The warp factor then reduces to
\begin{equation}\label{sol}
U^2(r)= \frac{ 3 \sqrt3 \left( \frac{4}{27} \xi_0 \xi_1^3 g^2 r^2+ 1 - \frac{\mu}{r}+ \frac{q_1^2}{\xi_1^2r^2}  \right)}{2 \sqrt{\xi_0 \xi_1^3} }\,.
\end{equation}
Solutions of this kind admit two noncoincident horizons for a suitable range of parameters, giving rise to nonextremal black holes.
When we set $\mu =0$ and we choose the positive sign in \eqref{constrscalars} we retrieve the supersymmetric solution with constant scalars found by \cite{sabra}\footnote{In particular the solution \eqref{sol} with $\mu =0$ is related through a redefinition of the coordinate $r$ to the BPS one in \cite{sabra}.}.
The solutions \eqref{sol} asymptote to ordinary AdS$_4$, with underlying $osp(2|4)$ superalgebra; for this reason the BPS bound is
\begin{equation}\label{BPSele}
M \geq |Z|\,,
\end{equation}
where $M$ is the mass of the configuration, and $Z$ is the charge related to the $U(1)\subset SU(2)_R$ symmetry generator in the $osp(2|4)$ superalgebra \footnote{In the conventions of \cite{kiril} this quantity is denoted as $T$.}.
Formulas and further details for these $AdS_4$ charges can be found in \cite{kiril}. In particular, the mass formula reads
$$
M= \frac{1}{8 \pi} \lim_{r \rightarrow \infty} \oint {\rm d}\Sigma_{tr} \bigg(e_{[0}^t e_{1}^r e_{2]}^{\theta} + \sin \theta e_{[0}^t e_{1}^r e_{3]}^{\varphi} +2g g' r | P_{\Lambda}^a L^{\Lambda}|e_{[0}^t e_{1]}^{r}+
$$
\begin{equation}\label{mass_ele}
  - \sqrt{g'^2 r^2 +1} (\omega_{\theta}^{12} e_{[0}^t e_1^{r} e_{2]}^{\theta} +\omega_{\varphi}^{13} e_{[0}^t e_1^{r} e_{3]}^{\varphi}) \bigg) \,,
\end{equation}
where $g'=g (\xi_0 \xi_1^3)^{1/4} \frac{\sqrt2}{3^{3/4}}$, and
\begin{equation}\label{central}
Z= \frac{1}{ 4 \pi}  \lim_{r \rightarrow \infty} \oint_{S^2} Re(T^-)=  \lim_{r \rightarrow \infty} Re(L^{\Lambda}q_{\Lambda}-M_{\Lambda} p^{\Lambda})\,.
\end{equation}
Computed on our solution \eqref{constrscalars}, \eqref{sol}, these formulas give 
\begin{equation}\label{M}
M= \frac{ (\xi_0 \xi_1^3)^{1/4} \mu}{3^{3/4} \sqrt2 }\,
\end{equation}
and (for positive sign in \eqref{constrscalars})
\begin{equation}\label{Z}
Z= \frac{(\xi_0 \xi_1^3)^{1/4} }{ 2 \sqrt2 \, 3^{3/4}} \left( \frac{q_0}{\xi_0} +3 \frac{q_1}{\xi_1} \right)= \frac{ \sqrt2 (\xi_0 \xi_1^3)^{1/4} }{ 3^{3/4}} \frac{q_1}{\xi_1}\,.
\end{equation}
One is now able to compute the extremality bound and compare it with the BPS bound $M \geq |Z|$. In particular, given the last two  formulas \eqref{M} \eqref{Z} we have
\begin{equation}\label{BPSness}
\mu_{BPS}= 2\, |\frac{q_1}{\xi_1}|\,.
\end{equation}
We can then compute the quantity $\Delta \mu$, namely the gap between extremality \eqref{extr} and BPS \eqref{BPSness} (for convenience we choose $q_1$ to be positive):
\begin{equation}
\Delta \mu= \mu_{extr}-\mu_{BPS}= 
 \frac{1}{ g |V_*|^{1/2}} \left[ \frac{\sqrt2}{3 }\left( \sqrt{1+4 Y^2}+2\right) \left( \sqrt{1+4 Y^2}-1 \right)^{1/2} - 2 Y \right]\ ,
\end{equation}
where we defined 
\begin{equation}
Y^2\equiv g^2 |V_*|\frac{q_1^2}{\xi_1^2}=-\frac{\Lambda}{3}\,\frac{q_1^2}{\xi_1^2}\ .
\end{equation}
In the last equation, we have used the value of the cosmological constant \eqref{cc}.

The mass gap $\Delta \mu$ is never negative. We compute this quantity in the limit of large AdS radius (small $\Lambda$), keeping the black hole charge $q_1$ comparable to the gravitino charge $e_1$ defined in \eqref{grav-charge}. Then, we have $Y<<1$, and we can expand
\begin{equation}
g^{-3}\Delta \mu = -\frac{4}{9}\Lambda \Big(\frac{q_1}{e_1}\Big)^3 + {\cal{O}}(\Lambda^2)\ .
\end{equation}

\subsubsection{Magnetic solution}

In this section we will deal with purely magnetic black holes. These solutions have a constant magnetic flux at infinity; we furthermore impose a Dirac-like quantization condition on the charges:
\begin{equation}
g \xi_{\Lambda} p^{\Lambda}= -1\,.
\end{equation}
In this way these states asymptote to the so-called magnetic AdS (mAdS) ground state. The concept of mAdS as a supersymmetric ground state of $\mathcal{N}=2$ gauged supergravity has been introduced in \cite{ours}, based on the supersymmetric state found for the first time in \cite{romans}. These ground states have a significantly different form of the Killing spinor and for this reason they are endowed with a different superalgebra, characterized by the BPS bound
\begin{equation}\label{BPSboundmagn}
M \geq 0 \,.
\end{equation}
More details about mAdS can be found in \cite{ours} and in Section 4. We will use is the mass formula for asymptotically mAdS solutions in presence of scalars \cite{kiril}, different from the mass formula in ordinary AdS without magnetic charge :
$$
M= \frac{1}{8 \pi} \lim_{r \rightarrow \infty} \oint {\rm d}\Sigma_{tr} \left( g' r+ \frac{g'}{2 g^2 r}\right)  \bigg( 2 {\text Im}(L^{\Lambda} q_{\Lambda} -M_{\Lambda}p^{\Lambda}) \sin \theta e_0^t e_1^r e_2^{\theta} e_3^{\varphi} +
$$
\begin{equation}\label{mass_mads}
+2g | P_{\Lambda}^a L^{\Lambda}|e_0^t e_1^{r}  - (\omega_{\theta}^{12} e_0^t e_1^{r} e_2^{\theta} +\omega_{\varphi}^{13} e_0^t e_1^{r} e_3^{\varphi}) \bigg)\,. 
\end{equation}
We choose again $b=0$, and in order to make contact with the solutions derived in Section 5, we take $a=\frac{3^{3/4} }{\sqrt{2} (\xi_0 \xi_1^3)^{1/4}} $, so that 
\begin{equation}
h(r)=\frac{3^{3/4}  r }{\sqrt{2} (\xi_0 \xi_1^3)^{1/4}}\,,\qquad
U^2(r)= \frac{{2 \sqrt{\xi_0 \xi_1^3}}  \left(g^2 r^2+ 1 - \frac{\mu}{r}+ \frac{Q}{r^2}  \right)} {3 \sqrt{3}}\,.
\end{equation}
Then \eqref{constr_2} imposes the constraint
\begin{equation}\label{eqQ}
Q= \frac{4 \xi_1^2 g^2 (p^1)^2+6 \xi_1 g p^1+3}{3 g^2}\,,
\end{equation}
while \eqref{constr_quadratic}  gives
\begin{equation}
8(p^1)^2 \xi_1^2 g^2+18p^1 \xi_1 g+9=0\,,
\end{equation}
that has two solutions:
\begin{equation}\label{solution1}
solution \,\,1 \qquad
p_{*}^1= -\frac{3}{4 g \xi_1}\,,  \quad p_{*}^0=-\frac{1}{4 g \xi_0}\,, \qquad \rightarrow \qquad Q=\frac{1}{4g^2}\,,
\end{equation}
\begin{equation}\label{strangep}
solution\,\, 2 \qquad p_{**}^1= -\frac{3}{2 g \xi_1}\,, \quad p_{**}^0=\frac{1}{2 g \xi_0}\,, \qquad  \rightarrow \qquad Q=\frac{1}{g^2}.
\end{equation}

For both these solutions the mass is
\begin{equation}
M= \frac{3^{3/4}\mu}{2 \sqrt2 (\xi_0 \xi_1^3)^{1/4}}\ .
\end{equation}

Let us examine more closely the two cases. For the case denoted by $p_{*}^1$ the warp factor is
\begin{equation}
{U^2}_{p=p_{*}^1}(r) = \frac{2 \sqrt{\xi_0 \xi_1^3}}{3 \sqrt3} \left(g^2 r^2 +1 - \frac{\mu}{r}+\frac{1}{4 g^2 r^2} \right)\,.
\end{equation}
For $\mu=0$ the warp factor is a perfect square, and supersymmetric solutions of this type include the $1/4$ BPS magnetic solution found in \cite{sabra}. These BPS solutions are singular. 

In the second case, $p_{**}^1$, instead, the warp factor is
\begin{equation}
{U^2}_{p=p_{**}^1}(r)=\frac{2 \sqrt{\xi_0 \xi_1^3}} {3 \sqrt3}\left(g^2 r^2 +1 - \frac{\mu}{r}+\frac{1}{g^2 r^2} \right)\ ,
\end{equation}
and when we set $\mu=0$ it also gives a naked singularity. Furthermore, the warp factor in this case it is not a perfect square. By studying the supersymmetry variations one can see that this solution is not supersymmetric in $\mathcal{N}=2$ gauged supergravity. Perhaps such a solution is supersymmetric in $\mathcal{N}=4$ or $\mathcal{N}=8$ supergravity, along the lines of \cite{Khuri:1995xk,Ferrara:2007pc,Bena:2011pi}.

At this point we can repeat here the procedure used in section to compute the mass gap $\Delta \mu$ between extremality and BPSness. Notice that in this case $\mu_{BPS}=0$. We take into consideration just the solution \eqref{solution1}, for which  $Q=\frac{1}{4g^2}$. In the limit of small cosmological constant, we find from \eqref{extr},
\begin{equation}
\Delta \mu = \mu_{extr}-\mu_{BPS}\approx \frac{1}{g}\Big(1+\frac{|V_*|}{2}\Big)\ .
\end{equation}

\section{Electric black holes with nonconstant scalars}

This section gives a detailed description of electrically charged black holes in the presence of nonconstant scalars. We mainly reproduce known results here \cite{Duff:1996hp}, but also some new solutions. They are deformations of the the supersymmetric ones found in \cite{sabra}, but unfortunately they all seem to contain naked singularities too. For the sake of completeness, we decided to keep them in our presentation. Perhaps these singularities can be resolved when adding higher derivative terms to the action.

\subsection{1/2 BPS electric solutions}

We start with describing the purely electric BPS solution of \cite{sabra}. Metric and symplectic sections for the BPS case are of the form \eqref{ansatz_metric} with
\begin{equation}
ds^2= e^{\mathcal{K}}(1+ g^2 r^2 e^{-2\mathcal{K}}) dt^2 -\frac{dr^2}{(1+ g^2 r^2 e^{-2\mathcal{K}})}-e^{-\mathcal{K}}r^2 d\Omega_2^2\,,
\end{equation}
\begin{equation}\label{harm_func}
{\text Im} X^{\Lambda}=0\,, \qquad
2 {\text Im} F_{\Lambda} =H_{\Lambda}= \xi_{\Lambda} + \frac{q_{\Lambda}}{r}\,.
\end{equation}

Again, we take the example of the prepotential $F=-2i \sqrt{X^0 (X^1)^3}$; the holomorphic sections then are\footnote{From now on we will use indistinctly two notations, namely with upper or lower indices for the sections $X^\Lambda$.}
\begin{equation}\label{sect}
X_0 = \frac{1}{6 \sqrt{3}} \sqrt{\frac{H_1^3}{H_0}}\,, \qquad X_1=\frac{1}{2 \sqrt{3}} \sqrt{H_0 H_1}\,,
\end{equation}
so that
\begin{equation}
e^{-\mathcal{K}}= \frac{2}{3 \sqrt3} \sqrt{H_0 H_1^3}\,.
\end{equation}
The solution is purely electric, with field strengths of the form (the sections are real, so by construction $R_{\Lambda \Sigma}=0$)
\begin{equation}
F_{tr}^{\Lambda}= \frac{I^{\Lambda \Sigma} q_{\Sigma}}{h^2(r)} \,,
\end{equation}
while the magnetic charges are set to zero $p^0=p^1=0$. For these BPS solutions the mass and the charge were computed in \cite{kiril} by means of the formulas \eqref{mass_ele} \eqref{central}, and give
\begin{equation}\label{cen}
M= \frac{(\xi_0 \xi_1^3)^{1/4} }{ 2 \sqrt2 \,3^{3/4}} \left( \frac{q_0}{\xi_0} +3 \frac{q_1}{\xi_1} \right)=Z\,.
\end{equation}
The BPS bound $M \geq |Z|$ is saturated by the BPS solutions, if we restrict to positive charges.

\subsection{Nonextremal electric solutions}
To find nonextremal solutions, we make the following modification to the ansatz for the warp factor:
\begin{equation}
U^2(r)=e^{\mathcal{K}}(1- \frac{\mu}{r}+ g^2 r^2 e^{-2\mathcal{K}})\,,
\end{equation}
keeping
\begin{equation}
h(r)=e^{-\mathcal{K}/2}r\,.
\end{equation}
The sections are computed from \eqref{sect}, where the harmonic functions $H_{\Lambda}$ are
\begin{equation}
H_{0}= a_0 +\frac{b_0}{r}\,, \qquad H_{1}=a_1 +\frac{b_1}{r}\,.
\end{equation}
At this moment $a_{\Lambda}$ and $b_{\Lambda}$ are unconstrained. We retrieve the BPS solution by choosing $a_{\Lambda}=\xi_{\Lambda}$ and $b_{\Lambda}=q_{\Lambda}$.

The relevant components of Einstein's equations to be satisfied are \eqref{einst1}, \eqref{einst2}, and \eqref{einst3} computed on the Special K\"ahler quantities given in Appendix. We choose to solve first the three Einstein's equations: for the configurations with running scalars, this is enough since the scalar equation of motion does not give further constraints (see the discussion at the end of Section 2).

The first equation \eqref{einst1} is automatically satisfied without imposing any other constraints on the sections, while the second one \eqref{einst2} imposes $a_{\Lambda}=\xi_{\Lambda}$. The remaining equation \eqref{einst3}, gives the two independent constraints:
\begin{equation}\label{constr1}
 (b_1^2\xi_0^2-b_0^2 \xi_1^2)(b_0^2 +b_0 \xi_0 \mu -q_0^2)=0\,,
\end{equation}
\begin{equation}\label{constr2}
3(b_1^2\xi_0^2-b_0^2 \xi_1^2)(b_1^2 +b_1 \xi_1 \mu -q_1^2)=0\,.
\end{equation}

At this point we have two choices. The first choice is to set $(b_0^2 +b_0 \xi_0 \mu -q_0^2)=0$ and $(b_1^2 +b_1 \xi_1 \mu -q_1^2)=0$. If we furthermore  parameterize
\begin{equation}\label{par0}
q_0= \xi_0 \mu \cosh Q_0 \sinh Q_0\,,
\end{equation}
\begin{equation}\label{par1}
q_1= \xi_1\mu \cosh Q_1 \sinh Q_1\,,
\end{equation}
with $Q_1$ and $Q_0$ real, we obtain two sets of solutions for $b_0$, $b_1$. The first set is
\begin{equation}\label{b_1electr}
b_1= \xi_1 \mu \sinh^2 Q_1 \,
\qquad
b_0= \xi_0 \mu \sinh^2 Q_0\,.
\end{equation}
This family of solutions was found in the literature by \cite{duff_liu} in the context of $\mathcal{N}=8$ gauged supergravity, as already mentioned. In the BPS limit, they become singular \cite{duff_liu}.

The second set of solutions,  keeping the parameterization \eqref{par0} \eqref{par1}, is this:
\begin{equation}
 b_1= -\xi_1 \mu \cosh^2 Q_1 \,,
\qquad
 b_0=-\xi_0 \mu \cosh^2 Q_0\,,
\end{equation}
and seems to present naked singularities. The BPS solution is retrieved if the take  the limit $\mu \rightarrow 0$ and $Q_{\Lambda} \rightarrow -\infty$.

We have a second choice when $(b_1^2\xi_0^2-b_0^2 \xi_1^2)=0$, namely the first term in the product of \eqref{constr1}, \eqref{constr2}:
\begin{equation}\label{pm}
b_1= \pm \frac{b_0 \xi_1}{\xi_0}\,.
\end{equation}
When we choose the plus ($+$) sign we have a solution with constant scalars. The equations of motion are satisfied if we impose these other two constraints:
\begin{equation}
q_0 = \pm \frac{\xi_0 q_1}{\xi_1}\,, \qquad 
\mu=\frac{  \xi_0^2 q_1^2 - b_0^2 \xi_1^2 }{ b_0 \xi_0 \xi_1^2}\,.
\end{equation}
The solution found represents genuine black hole for a suitable choice of parameters. Setting $\mu =0$ in the case of the plus ($+$) sign we retrieve the BPS solution of \cite{sabra} with constant scalars. This solution, in the notation of \eqref{const_b} has $b \neq 0$, nevertheless the scalar is constant due to the fact that sections are proportional to each others. 

If we instead choose the minus sign in \eqref{pm} we get new constraints but all the solutions seem to have naked singulaties.

\subsubsection{Mass and other charges of the configurations}

We make use of the formulae in order to find the charges of the nonextremal solutions. The charge $Z$ has the same value as in \eqref{cen}, 
\begin{equation}
Z= \frac{(\xi_0 \xi_1^3)^{1/4} }{ 2 \sqrt2 \, 3^{3/4}} \left( \frac{q_0}{\xi_0} +3 \frac{q_1}{\xi_1} \right)\,.
\end{equation}
The mass is computed from \eqref{mass_ele}; for all the nonextremal solutions found in the previous subsection it turns out to be
\begin{equation}
M= \frac{3 b_1 \left(\frac{\xi_0}{\xi_1} \right)^{1/4} + b_0 \left(\frac{\xi_1}{\xi_0} \right)^{3/4}+2  (\xi_0 \xi_1^3)^{1/4} \mu}{2\sqrt2 \, 3^{3/4}} \,.
\end{equation}
In the particular case of the solution \eqref{b_1electr}, it gives
\begin{equation}\label{mass_liu}
M= \frac{\mu (\xi_0 \xi_1^3)^{1/4}  (2+ 3 \sinh^2Q_1 +\sinh^2Q_0)}{2\sqrt2 \, 3^{3/4}} \,,
\end{equation}
in agreement with the result of \cite{Anabalon:2012sn} (apart from numerical factors due to normalization).

The condition for the saturation of the BPS bound $M=|Z|$ is (we assume $Z$ to be positive for simplicity):
\begin{equation}\label{saturationBPS}
2 \mu= \frac{q_0-b_0}{\xi_0}+3\frac{q_1-b_1}{\xi_1}\,.
\end{equation}
We recover the supersymmetric solution by imposing $\mu=0$, and $q_{\Lambda}=b_{\Lambda}$; but there are also in this case other solutions that saturate the bound, namely those for which \eqref{saturationBPS} in satisfied with $\mu \neq 0$.

\section{Magnetic black holes with nonconstant scalars}

In this section we find nonextremal magnetic black holes with nonconstant scalars. They are a deformation of the $1/4$ BPS solution found in \cite{kiril_stefan}. Before describing the deformation procedure, we briefly recap the features of the undeformed BPS solution  found in \cite{kiril_stefan} based on \cite{klemm-adsBH}. 

\subsection{1/4 BPS magnetic black hole solution}
The supersymmetric solution was found in nonminimal gauged supergravity with one vector multiplet, in presence of prepotential $F= -2i \sqrt{X_0 X_1^3}$. It is an extremal BPS black hole with vanishing electric field strengths
\begin{equation}
F_{tr}^0=F_{tr}^1=0 \,,
\end{equation}
and a real scalar.  $R_{\Lambda \Sigma}$ vanishes in this case, so then for consistency the electric charges are set to zero:
\begin{equation}
q_{0}=q_1=0\,.
\end{equation}
The magnetic field strengths are of the form \eqref{ansatz_fs} with magnetic charges obeying the Dirac-like quantization condition\footnote{There exists also another solution with $g \xi_{\Lambda}p^{\Lambda}= +1$ that can be treated in all similarity.}
\begin{equation}\label{quant}
g \xi_{\Lambda}p^{\Lambda}= -1\,.
\end{equation}
The static and spherically symmetric metric is of the type \eqref{ansatz_metric}, where 
\begin{equation}
U^2(r)= e^{\mathcal{K}}\left(gr+\frac{c}{2gr} \right)^2\,,
\end{equation}
and
\begin{equation}\label{same}
h(r)= r e^{-\mathcal{K}/2}\,.
\end{equation}
In particular, running scalars allow for a negative parameter $c$, so that the solution can have a horizon. The sections are harmonic functions:
\begin{equation}
z= \frac{X_1}{X_0}\,, \qquad X_0 = a_0 + \frac{b_0}{r}\,, \qquad X_1 = a_1 + \frac{b_1}{r}\,,
\end{equation}
whose parameters are constrained by the Killing spinor equations to be
\begin{equation}\label{param_BPS}
a_0= \frac{1}{4 \xi_0}\,, \qquad b_0= -\frac{\xi_1 b_1}{\xi_0}\,, \qquad a_1= \frac{3}{4 \xi_1}\,, \qquad
c=1-\frac{32}{3}(g \xi_1 b_1)^2\,.
\end{equation}
In particular also the magnetic charges are fixed:
\begin{equation}\label{magnetic}
p^0= - \frac{2}{g \xi_0} \left(\frac18 +\frac{8 (g \xi_1 b_1)^2}{3} \right)\,,\,\,\,\,\,\,\,\,\,\,\,\,\,
p^1=- \frac{2}{g \xi_1} \left(\frac38 -\frac{8 (g \xi_1 b_1)^2}{3} \right)\,,
\end{equation}
or, vice versa, $b_1$ can be written in function of the magnetic charges.

The singularity is not situated at $r = 0$, and this point is not even a horizon.
Genuine singularities will appear at $r_{s} =  4 \xi_1 b_1 , - \frac43 \xi_1 b_1$. The horizon, instead, is at
\begin{equation}
r_h = \sqrt{\frac{16}{3}(\xi_1 b_1)^2 -\frac{1}{2g^2}}\ .
\end{equation} 
The requirement that the horizon must shield the singularity,  namely $r_h>r_{s}$, sets some constraints on the value of $g \xi_1 b_1$.
To sum up, the BPS  solution found represents a black hole in a particular range of parameters. A detailed description of this and other properties of the solution can be found in the original paper \cite{kiril_stefan}.

\subsection{Nonextremal magnetic solution}
We are now ready to deform the BPS solution of the previous section to a nonextremal black hole. We use the minimal modification ansatz for the warp factor:
\begin{equation}\label{nonextr_new_sol}
U^2(r)= e^{\mathcal{K}}\left(g^2 r^2+c - \frac{\mu}{r}+\frac{Q}{r^2} \right)\,,
\end{equation}
together with \eqref{same}. The scalar $z$ is real and positive and we keep the same sections as in the BPS case, 
\begin{equation}
z= \frac{X_1}{X_0}\, \qquad
X_0 = a_0 + \frac{b_0}{r}\,, \qquad  X_1 = a_1 + \frac{b_1}{r}\,.
\end{equation}
This guess for the form of the nonextremal solution is then followed by brute-force solving the Einstein's equations of motion. Indeed the first equation \eqref{einst1} is automatically satisfied given the form of the sections. For \eqref{einst2} to be satisfied, instead, we have to impose the same relations \eqref{param_BPS} as for the BPS case,
meanwhile solving the last equation \eqref{einst3} requires: 
\begin{equation}\label{c2}
Q= \left(-\frac{16}{3} b_1^2 \xi_1^2 + \frac{1}{g^2} -\frac{256}{27} b_1^4 \xi_1^4 g^2 + \frac{2 \xi_1 p^1}{g} +\frac{4}{3} \xi_1^2 (p^1)^2\right)\,,
\end{equation}
and
\begin{equation}\label{mu}
b_1 \mu= \left(\frac{8}{3} b_1^2 \xi_1 - \frac{3}{4 g^2 \xi_1} +\frac{512}{27} b_1^4 \xi_1^3 g^2 - \frac{3 p^1}{2g} -\frac{2}{3} \xi_1(p^1)^2\right)\,.
\end{equation}
The scalar equations of motion does not impose further constraints.
We can now make connection with the constant scalar solutions of Section 3.1.2. Indeed if we impose $b_1=0$ the sections become constant. Then the last equation leaves $\mu$ unconstrained, as we found in 3.1.2, provided that we impose the values \eqref{solution1} \eqref{strangep} for the magnetic charges. This is the remnant of the further constraint that one has to solve in the case of constant scalars.

In deriving \eqref{c2} \eqref{mu} we already used the fact that the magnetic charges should satisfy the quantization condition \eqref{quant}, since we want our state to be asymptotically mAdS. In this way we eliminated the dependence of the warp factor from the magnetic charge $p^0$.

To sum up, our solution depends on the FI parameters $\xi_1$, $\xi_0$ and on other two unconstrained parameters: $b_1$, $p^1$. In contrast, the BPS solution of \cite{kiril_stefan} depends just on the FI and on $p^1$: the nonextremal solution has one parameter more, $b_1$, that is considered for our purposes the non-extremality parameter. 

Plugging into the value of the charge \eqref{magnetic} we retrieve indeed the BPS black hole of \cite{kiril_stefan}: the solution becomes extremal.
We can say then that we have found a one-parameter nonextremal generalization of the magnetic solution found in \cite{kiril_stefan}, that by construction at extremality remains regular (no naked singularities).
Expressing the non-extremality in terms of $\mu$ seems more natural and it would in principle be possible to write $b_1$ in function of $\mu$ given the relation \eqref{mu}.

For completeness we give the value of the prefactor $e^{\mathcal{K}}$
\begin{equation}\label{ekappa}
e^{\mathcal{K}}= \frac{1}{8 \sqrt{ \left(\frac{3}{4 \xi_1}+\frac{b_1}{r} \right)^3 \left(\frac{1}{4 \xi_0}-\frac{b_1 \xi_1}{\xi_0 r} \right)}}
\end{equation}
so, together with \eqref{param_BPS}, \eqref{c2} and \eqref{mu} one has all the necessary values in order to retrieve the full warp factor $U^2(r)$.
We omit the full lengthy writing of it, nevertheless it is clear that the singularities lie at the zeros of formula \eqref{ekappa}
\begin{equation}\label{r-sing}
r_{s,1} =  4 \xi_1 b_1\, , \qquad r_{s,2}= - \frac43 \xi_1 b_1\,.
\end{equation}
These nonextremal black holes share with the BPS solution described in the previous section the feature of having a singularity at finite nonzero $r$. To have a proper black hole the  singularity must be shielded by a horizon, namely $r_{s}<r_{hor}$, and again this can be achieved by a suitable choice of the parameters of the solution.

\subsubsection{Mass of the magnetic nonextremal black hole}
We make use of the formula \eqref{mass_mads} to compute the mass for the new nonextremal magnetic solution. It turns out to be
\begin{equation}\label{result_mass}
M =- \frac{ \big( -9 +2 \xi_1 g (8 b_1^2 \xi_1 g -3 p^1) \big)  \big( -9+4 \xi_1 g (16 b_1^2 \xi_1g -3p^1) \big) }{ 72  \sqrt{2} \,3^{1/4}  b_1 (\xi_0 \xi_1^7)^{1/4}g^2  } \,.
\end{equation}
We have checked in numerous cases that the mass is positive in the absence of naked singularities.

The mass formula in \eqref{result_mass} admits two zeros: one is in correspondence with the value assumed by $p^1$ for the solution of \cite{kiril_stefan}
\begin{equation}
p^1= -\frac{2}{\xi_1 g} \left( \frac38- \frac83 (g \xi_1 b_1)^2 \right) \,.
\end{equation}
We retrieved then the expected result, namely that the supersymmetric solution saturates the BPS bound given by \eqref{BPSboundmagn}. 
We have then another zero of the mass formula obtained for
\begin{equation}\label{strange}
p^1= -\frac{2}{\xi_1 g} \left(\frac34 -\frac43 (g \xi_1 b_1)^2 \right)\,.
\end{equation}
This value does not seem to correspond to any BPS solution found so far in $\mathcal{N}=2$ gauged supergravity. The warp factor in this case is not a square, and all solutions of this family correspond to naked singularities. As we already mentioned, it could be that these solutions are supersymmetric in other models of extended supergravity \cite{Khuri:1995xk,Ferrara:2007pc,Bena:2011pi}.

\section{Product of the areas}

With the nonextremal solution at hand we can check a conjecture about the area product formula of black hole horizons \cite{Larsen:1997ge}. It seems true in a lot of examples \cite{Cvetic:2010mn,Galli:2011fq, Castro:2012av} that the product of the areas of the inner and outer horizons of a black hole is just function of the quantized charges (and, if present, of the cosmological constant), in particular it is independent of the mass or any other continuous moduli parameters. Such a product area law might be a calling for an underlying microscopic interpretation in string theory or conformal field theory. 

As discussed in \cite{Cvetic:2010mn}, for AdS$_4$ the product area law holds if we take into consideration all the roots of the warp factor, including the zeroes at complex values of $r$. A physical meaning of this is still not clear, nevertheless we can check the product area law for our solutions.

We deal with constant scalar solutions first. For these solutions  (section 3.1.1 and 3.1.2), the warp factor \eqref{formU} ($b=0$) can be decomposed in this way:
\begin{equation}
r^2 U^2(r)=\frac{ g^2 |V_{*}|}{3r^2} \prod_{\alpha=1}^4 (r-r_{\alpha})\,.
\end{equation}
The product of the four roots $r_1 r_2 r_3 r_4$, then, is the coefficient of lowest degree in $r$, namely it is proportional to the quantity denoted with $Q$ in \eqref{formU}.

For the electric black hole, then, using \eqref{Qel} together with \eqref{constrscalars}, we have\footnote{Notice that we have to compute the area of the horizons using the metric component $h(r)$ in our coordinates.}:
\begin{equation}
 \prod_{\alpha=1}^4 A_{\alpha}= (4 \pi)^4 \prod_{\alpha=1}^4 h(r_{\alpha})^2= (4 \pi)^4 \frac{q_1^4}{(g \xi_1)^4}= (4 \pi)^4 \frac{q_1^4}{e_1^4}\,.
\end{equation}
We recall that the quantities $e_{\Lambda}= g \xi_{\Lambda}$ are the electric charges of the gravitinos. 

For the magnetic black hole, instead, from \eqref{eqQ} and \eqref{solution1}, the product reads
 \begin{equation}
 \prod_{\alpha=1}^4 A_{\alpha} = (4 \pi)^4 \prod_{\alpha=1}^4 h(r_{\alpha})^2=  (4 \pi)^4 \frac{3^6}{2^8(e_0 e_1^3)^2}= (4 \pi)^4 \frac{27 p^0 (p^1)^3}{e_0 e_1^3} \,,
\end{equation}
For both constant scalar cases, the product does not depend on the mass parameter $\mu$. 

Things get more involved when one takes into account running scalars. In both electric and magnetic cases (for the electric black holes we refer in particular to formulas \eqref{par0}, \eqref{par1} \eqref{b_1electr}) the product of the areas is:
\begin{equation}\label{prodkappa}
 \prod_{{\alpha}=1}^4 A_{\alpha} = (4 \pi)^4 \prod_{\alpha=1}^4 h(r_{\alpha})^2=(4 \pi)^4 \prod_{\alpha=1}^4 e^{-\mathcal{K}(r_{\alpha})} r_{\alpha}^2 \,,
\end{equation} 
where the function $h^2(r)$ is of the form
\begin{equation}\label{acca}
h^2(r)= {\rm const} \times \sqrt{(r-r_{s,1})(r-r_{s,2})^3}\,,
\end{equation}
with $r_{s,1/2}$ the location of the singularities.
In particular, for the magnetic case $r_{s,1/2}$ are given by \eqref{r-sing}.
We now rewrite the warp factor as:
\begin{equation}
U^2(r) =  {\rm const'} \times \frac{e^{\mathcal{K}}}{r^2} \prod_{\alpha=1}^4 (r-r_{\alpha})\,.
\end{equation}
For instance, in the magnetic case we have:
\begin{equation}
U^2(r)=e^{\mathcal{K}} \left(g^2 r^2 +c- \frac{\mu}{r} +\frac{Q}{r^2} \right)=\frac{e^{\mathcal{K}}g^2}{r^2} \left(r^4 +\frac{c r^2}{g^2}- \frac{\mu r}{g^2} +\frac{Q}{g^2} \right) = \frac{e^{\mathcal{K}}g^2}{r^2} \prod_{\alpha=1}^4 (r-r_{\alpha})\,.
\end{equation}
The coefficient of lowest degree in $r$, namely $\frac{Q}{g^2}$, gives the value the product of all the roots $r_1 r_2 r_3 r_4$.
We now first make the redefinition
\begin{equation}
r'=r-r_{s,1}\,,
\end{equation}
and we express the warp factor in terms of $r'$. In a similar way the coefficient of lowest degree in $r'$, from now on denoted by $\kappa_1$, represents the product of all the $r'$ roots: $r'_1 r'_2 r'_3 r'_4$. For instance, in the magnetic case this coefficient turns of to be:
\begin{equation}\label{kappa-magn}
\kappa_1= r_{s,1}^4 +\frac{c r_{s,1}^2 }{g^2}-\frac{\mu  r_{s,1}}{g^2}+\frac{Q}{g^2}\,,
\end{equation}
where the values of $\mu$ and $Q$ are given respectively in \eqref{mu} \eqref{c2}. Repeating the procedure for $r_{s,2}$ gives $\kappa_2$, so that we have what we need to compute the area product. Using \eqref{acca} and \eqref{prodkappa}, we arrive at
\begin{equation}
 \prod_{\alpha=1}^4 A_{\alpha}= ({\rm const})^4 (4 \pi)^4 \sqrt{ \kappa_1 \kappa_2^3}\,.
\end{equation}
This formula holds for both electric and magnetic solutions, with the appropriate values of $\kappa_1$ and $\kappa_2$.

Similar to \eqref{kappa-magn}, we can compute $\kappa_1$ and $\kappa_2$ for the electric solution. We refrain from giving explicit formulas, since they are rather lengthy. An explicit calculation of the product area law for the electric solution yields\footnote{We took positive charges in \eqref{prod-area} and \eqref{prod-area2}. For negative charges, one must take absolute values.}
\begin{equation}\label{prod-area}
 \prod_{\alpha=1}^4 A_{\alpha} = \prod_{\alpha=1}^4 e^{-\mathcal{K}(r_{\alpha})} r_{\alpha}^2 = (4\pi)^4 \frac{q_0 q_1^3}{e_0 e_1^3} \,.
\end{equation}
We see then that the result  depends only on the black hole and gravitino electric charges. This result agrees with (the static limit of) the one of \cite{Cvetic:2010mn}.

For the magnetic case we find
\begin{equation}\label{prod-area2}
 \prod_{\alpha=1}^4 A_{\alpha}  = \frac{ (4\pi)^4 3^3}{g^4} \left(\frac{1}{g^2} +\frac{2 p^1}{g}+(p^1)^2 \right)^{1/2} (p^1)^3=  27 (4\pi)^4\,\frac{(p^0) (p^1)^3}{e_0e_1^3} \ .
\end{equation}
Also this product is  function solely of the magnetic charges $p^{\Lambda}$ of the black hole, and the electric charges of the gravitinos.

A few comments are in order here: first of all, we can see that the product area formulae obtained for constant and nonconstant scalar solutions are identical. Secondly, the dependence on the charges resembles the form of the prepotential of the model: it would be interesting to find any direct explanation of this. Finally we rewrite more suggestively the formulas in the electric and magnetic case in terms of the cosmological constant $\Lambda = 3 g^2 V_{*}$ (see \eqref{vstar} and \eqref{cc}) and the charges:
\begin{equation}
\prod_{\alpha=1}^4 A_{\alpha, electric}=(4\pi)^4 \frac{12}{\Lambda^2}\,q_0q_1^3\,, \quad \qquad \prod_{\alpha=1}^4 A_{\alpha, magnetic}= (4\pi)^4\frac{ 3^4 2^2 }{\Lambda^2 }(p^0) (p^1)^3 \,.
\end{equation}
In this way, the electric charges of the gravitinos are traded for the cosmological constant, and only the black hole charges appear explicitly.

\section{Magnetic black brane}

As a last example, here we illustrate that the same minimal modification ansatz used for the black hole works also in generating the nonextremal deformation of the black brane solution found in \cite{klemm-adsBH,brane}. Black branes in gauged supergravity were also studied in e.g. \cite{Barisch:2011ui} and particular aspects of magnetic black branes were analyzed in detail  in \cite{Donos:2011qt}.
The magnetic 1/4 BPS black brane given in \cite{brane} is characterized by this metric:
 \begin{equation}\label{our_solution}
 {\rm d} s^2 = U^2(r) \,{\rm d}t^2 - \frac{ {\rm d}r^2}{U^2(r)} -h^2(r)\, {\rm d} \sigma^2\ ,
\end{equation}
where
\begin{equation}
U^2(r)=e^{\mathcal{K}} \left(g r+\frac{c}{2 g r} \right)^2 \,,
\qquad
h(r)=e^{-\mathcal{K}/2}r\,,
\end{equation}
and the area element is given by 
\begin{equation}
{\rm d}\sigma^2= \frac{\mathcal{V}}{{\rm Im}\tau}({\rm d}x^2 +2 {\rm Re}\tau {\rm d}x {\rm d}y+ {\rm d}y^2)\,.
\end{equation}
The field strengths have this form:
\begin{equation}
F_{rt}=0\,, \qquad F_{xy}^{\Lambda}=\frac{ p^{\Lambda}}{2} \mathcal{V}\,,
\end{equation}
and the magnetic charges satisfy this relation, found by \cite{klemm-adsBH}\footnote{Moreover, in \cite{klemm-adsBH} was found a relation between the integer number on the RHS of the Dirac-like quantization condition and the curvature of the event horizon: $g \xi_{\Lambda}p^{\Lambda} =k$, with $k= -1,0,1$.}:
\begin{equation}\label{dirac-brane}
\xi_{\Lambda} p^{\Lambda} = 0\,.
\end{equation}

In what follows we will restrict ourselves to the simple case of rectangular torus,  $ {\rm Im} \tau =1$.
The explicit example of the BPS black brane is again given for the  prepotential $F =-2 i \sqrt{X^0 (X^1)^3}$. We have $X^0 = H^0 =a_0+\frac{b_0}{r}, X^1 = H^1 = a_1+\frac{b_1}{r}$ and $e^{-\mathcal{K}} = 8 \sqrt{H^0 (H^1)^3}$, with
 \begin{equation}\label{BPSbrane}
b_0 = -\frac{\xi_1 b_1}{\xi_0}\,, \qquad a_0 = -\frac{1}{4 \xi_0}\,, \qquad a_1 = -\frac{3}{4 \xi_1}\,, \qquad c = - \frac{32}{3} (g \xi_1 b_1)^2\,,
\end{equation}
and magnetic charges
\begin{equation}\label{magn-charges}
p^0 = \frac{16 (g \xi_1 b_1)^2}{3 g \xi_0}\,, \quad p^1 = -\frac{16 (g \xi_1 b_1)^2}{3 g \xi_1}\,.
\end{equation}

We now want to generalize the ansatz for $U(r)$, in order to find a solution with two noncoincident horizon. We furthermore keep the Dirac quantization condition \eqref{dirac-brane}.
The relevant equations to be satisfied in this case are given as in the black hole case by some linear combinations of the Einstein's equations, namely:\\
adding $tt$ and $rr$:
\begin{equation}
-2 \frac{h''}{h}=\frac{3}{8z^2}\partial_{r} z \partial_{r} z\,,
\end{equation}
adding $rr$ and $\theta \theta$:
\begin{equation}
-\frac{(U^2\,h^2)''/2}{h^2}  =  - 2g^2 \left(\frac{\xi_0 \xi_1}{\sqrt{z}}+ \frac{\xi_1^2}{3} \sqrt{z}\right)\,,
\end{equation}
subtracting $tt$ from $\theta \theta$:
\begin{equation}\label{third_brane}
U^2 h'^2 +h h'' U^2 -h^2 U'^2-h^2 U U'' =- \frac{1}{h^2} \left( (p^{0})^2 \sqrt{z^3} +  \frac{3(p^{1})^2}{\sqrt{z}} \right)\,,
\end{equation}
If these are satisfied for running scalars, the scalar equation of motion does not impose any further constraints.

Having faith in our usual minimal modification ansatz, we propose this form of solution:
\begin{equation}\label{branee}
U^2(r)= e^{\mathcal{K}}\left(g^2 r^2 +c - \frac{\mu }{r} + \frac{Q}{r^2} \right)\,, \qquad h(r)=e^{-\mathcal{K}/2}r\,,
\end{equation}
keeping the same form of the scalars:
\begin{equation}
z= \frac{X_1}{X_0} \,, \qquad X_1= a_1 +\frac{b_1}{r}\,, \qquad  X_0= a_0+\frac{b_0}{r}\,,
\end{equation}
with the same relations \eqref{BPSbrane} as in the BPS case. The parameters appearing in the warp factor are constrained in this way by the equation \eqref{third_brane}:
\begin{equation}
\mu = -\frac{512}{27} g^2 \xi_1^3 b_1^3 +\frac{2}{3}\frac{\xi_1 (p^1)^2}{b_1}\,,
\end{equation}
\begin{equation}\label{Qbrane}
Q = - \frac{256}{27} g^2 \xi_1^4 b_1^4 +\frac43 \xi_1^2 (p^1)^2\,.
\end{equation}
These solutions satisfies all the equations of motion, also the scalar one. 

The nonextremal black brane just found shares some feature with the nonextremal black hole of Section 4.2. For instance, the singularity is at finite nonzero $r$, and for a suitable choice of parameters there are two noncoincident horizons shielding the singularity. Furthermore, also in this case  the family of solutions depends on the four parameters $\xi_0$, $\xi_1$, $p^1$, $b_1$. By choosing the value \eqref{magn-charges} the solution reduces to the extremal 1/4 BPS one of \cite{kiril_stefan}.

\subsection{Mass and charge of the black brane}

Based on the analysis done in \cite{brane}, the underlying superalgebra in the case of the magnetic black brane is characterized by the BPS bound $M \geq |Z|$, where $M$ is the mass of the configuration and  $Z$ is a (real) central charge. These two quantities are respectively given by:
\begin{equation}\label{mass}
M=\frac{\mathcal{V}}{2} \lim_{r\to \infty}{\oint {\rm d} \Sigma_{tr} e_{0}^{t} e_{1}^{r} \left(\, 2 g r |P^a_{\Lambda} L^{\Lambda}| -r (\omega_{x}^{1 2} e_{2}^{x} + \omega_{y}^{13} e_{3}^{y} + \omega_{y}^{12} e_{2}^{y} ) \,  \right)}
\end{equation}
and
\begin{equation}\label{ccharge}
Z= 2 \mathcal{V} \lim_{r \rightarrow \infty} \oint_{T^2} r\ {\rm Im} \left( T^- \right) = 2 \lim_{r \rightarrow \infty} r \mathcal{V}\ {\rm Im}\left( L^{\Lambda} q_{\Lambda} - M_{\Lambda} p^{\Lambda} \right)\ ,
\end{equation}
where $T^-$ is the anti-selfdual part of the graviphoton field strength. 

Contrary to their spherical analogs, the magnetic BPS black branes have a non-vanishing mass. 
Furthermore, to have the correct asymptotic behaviour (namely to admit an asymptotic Killing spinor) the solution must satisfy ${\rm lim}_{r \rightarrow \infty}{\rm Im} ( L^{\Lambda}q_{\Lambda} - M_{\Lambda}p^{\Lambda})=0$, which one can verify explicitly. See \cite{brane} for further details. 

The mass of the nonextremal configuration \eqref{branee} \eqref{third_brane} \eqref{Qbrane} can be computed from
\eqref{mass}, and the result is 
\begin{equation}
M=  \mathcal{V} \left( \frac{128 g^3 b_1^4 \xi_1^3 +9 g \xi_1 (p^1)^2}{27 b_1 g^2} \right)\,,
\end{equation}
while the central charge formula gives
\begin{equation}
Z= -\mathcal{V} \left( \frac{8 \xi_1^2 b_1 p^1}{3} \right)\,.
\end{equation}
For the cases when a proper black brane exist (no naked singularities), the BPS bound $M\geq |Z|$ is satisfied.
If we search for the values of the parameters for which the bound is saturated, we find that there are two of them:
\begin{equation}
p^1 = -\frac{16 (g \xi_1 b_1)^2}{3 g \xi_1} \qquad p^1 = -\frac{8 (g \xi_1 b_1)^2}{3 g \xi_1} \ .
\end{equation}
The first one corresponds to the BPS solution of \cite{klemm-adsBH,brane}. The other one, instead, is similar to the one we have in the black hole case (cfr formula \eqref{strange}): it does not preserve any supersymmetry within $\mathcal{N}=2$ gauged supergravity.

\section*{Acknowledgements}

We would like to thank A. Gnecchi, S. Minwalla, T. Ort\'{\i}n, M. J. Rodriguez, and especially K. Hristov for useful comments and discussions. We acknowledge support by the Netherlands Organization
for Scientific Research (NWO) under the VICI grant 680-47-603.

\section*{Appendix 1: Conventions and special K\"ahler quantities}

In our conventions the signature is $[+,-,-,-]$ and Riemann-Christoffel tensor and the Ricci tensor are defined as
\begin{equation}
R^{\rho}_{\sigma \mu \nu} = -(\partial_{\mu} \Gamma^{\rho}_{\nu \sigma}- \partial_{\mu} \Gamma^{\rho}_{\nu \sigma} + \Gamma^{\rho}_{\mu \lambda}\Gamma^{\lambda}_{\nu \sigma} - \Gamma^{\rho}_{\nu \lambda}\Gamma^{\lambda}_{\mu \sigma})\,, \qquad
R^{\rho}_{\sigma \rho \nu} = R_{\sigma \nu}\,.
\end{equation}
The Einstein's equation then read:
\begin{equation}
-(R_{ \mu \nu}- \frac12 R g_{\mu \nu})= \kappa^2 T_{\mu \nu}\,,
\end{equation}
where $T_{00}$ is positive. We furthermore take $\kappa =1$.

The Levi-Civita tensor is defined in this way:
\begin{equation}
\epsilon^{0123}=1=-\epsilon_{0123}\,,
\end{equation}
with
\begin{equation}
\epsilon^{\mu \nu \rho \sigma} = \sqrt{-det\, g} \, \, {{e}^{\mu}}_{a} {{e}^{\nu}}_b {{e}^{\rho}}_c {{e}^{\sigma}}_d \epsilon^{abcd}\,.
\end{equation}

Our conventions on special K\"ahler quantities are as in \cite{abcd}. We restrict ourselves to a model with prepotential $F(X)= -2i \sqrt{X^0 ({X^1})^3}$, just one scalar field. 
The relevant special K\"ahler quantities read:
\begin{equation}
\mathcal{K}=-log[i(\overline{X}^{\Lambda}F_{\Lambda} - \overline{F}_{\Lambda}X^{\Lambda})]=-log[X^0 \overline{X}^0 (\sqrt{z}+\sqrt{\bar{z}})^3]\,,
\end{equation}
such that the K\"ahler metric is
\begin{equation}
g_{z \bar{z}}= \partial_{z} \partial_{\bar{z}} \mathcal{K}=\frac34 \frac{1}{(\sqrt{z}+\sqrt{\bar{z}})^2 \sqrt{z \bar{z}}}\,.
\end{equation}
In the case in which $Im(z)=0$ we have to require $z$ positive if we want the metric to be positive definite.
The scalar potential turns out to be
$$
V(z, \bar{z})=(g^{i \bar{\jmath}} f_i^{\Lambda}\bar{f}_{\bar{\jmath}}^{\Sigma}-3\bar{L}^{\Lambda}L^{\Sigma}) \xi_{\Lambda} \xi_{\Sigma} = 
$$
\begin{equation}
 = \frac{1}{(\sqrt{z}+ \sqrt{\bar{z}})^3} \left( \left(-2 z - 2 \bar{z} -4 \sqrt{z \bar{z}}\right)\xi_0 \xi_1 - \left(\frac43 z \bar{z} +\frac23 (z +\bar{z}) \sqrt{z \bar{z}}\right)\xi_1^2 \right)\,,
\end{equation}

and the period matrix is
\begin{equation}
\mathcal{N}_{00}= i \frac{2\sqrt{z^3} \sqrt{\bar{z}}}{\sqrt{\bar{z}} -3 \sqrt{z}}\,,\,\,\,\,\,\,\,\,
\mathcal{N}_{01}= i\frac{3 (z - \sqrt{z \bar{z}})}{\sqrt{\bar{z}}-3 \sqrt{z}}\,,\,\,\,\,\,\,\,\,\,
\mathcal{N}_{11}= i \frac{6}{\sqrt{\bar{z}}-3 \sqrt{z}}\,.
\end{equation}



It is useful to compute the determinant of the real part of the period matrix, $R_{\Lambda \Sigma}$:
\begin{equation}
{\text Det[}R_{\Lambda \Sigma}] = \frac{3}{4} \frac{(\sqrt{z} -\sqrt{\bar{z}})^2 (z - 14 \sqrt{z \bar{z}} +\bar{z})}{(3z-10 \sqrt{z \bar{z}}+3 \bar{z})}\,.
\end{equation}
$R_{\Lambda \Sigma}$ has zero eigenvalues if and only if we take a real scalar. In order for it to be in the K\"ahler cone we assume it to be  positive.

With these assumptions, that are the ones we stick to in the explicit examples, the K\"ahler metric reduces to 
\begin{equation}
g_{z \bar{z}}= \frac{3}{16 z^2}\,.
\end{equation}
Given the real sections the period matrix is purely imaginary and diagonal:
\begin{equation}
\mathcal{N}_{\Lambda \Sigma}=
\left(
\begin{array}{cc}
-i \sqrt{z^3} & 0 \\
0 & -3i \sqrt{\frac{1}{z}} \\
\end{array}
\right)\,,
\end{equation}
and the scalar potential reduces to:
\begin{equation}
V(z, \overline{z})= -g^2 \left(\frac{\xi_0 \xi_1}{\sqrt{z}}+ \frac{\xi_1^2}{3} \sqrt{z}\right)\,.
\end{equation}

\section*{Appendix 2: Temperature}

Given that our metric is of the form \eqref{ansatz_metric},
\begin{equation}
{\rm d}s^2=U^2(r) {\rm d}t^2- \frac{1}{U^2(r)} {\rm d}r^2-h^2(r) {\rm d} \Omega^2\,,
\end{equation}
we denote the position of the outer and inner horizon with  $r_{+}, r_{-}$ respectively. Then
with a new radial coordinate $\rho= \sqrt{2 \kappa^{-1}|r-r_{+}|}$ we can write the near horizon $r\rightarrow r_{+}$ metric as
\begin{equation}\label{metric_temp}
{\rm d} s^2={\rm d} \rho^2+ \rho^2 (i \kappa {\rm d}t)^2 +h^2(r_{+}) {\rm d} \Omega^2 \,,
\end{equation}
where the surface gravity $\kappa$ is defined as
\begin{equation}
\kappa= \frac12 |2 U(r) U(r)'|_{r=r_{+}}\,.
\end{equation}
The metric \eqref{metric_temp} has a regular point for $\rho=0$ if $i \, t \kappa$ has period $2 \pi$, namely if $i \, t$ is an angular variable with period $2 \pi / \kappa$. The Hawking temperature at the horizon is given by the inverse of this period:
\begin{equation}
T= \frac{\kappa}{2 \pi}\,.
\end{equation}
Specializing to the magnetic black hole solution \eqref{nonextr_new_sol} of Section 5.2, we have:
\begin{equation}\label{temp}
T=\frac{1}{4 \pi} \bigg{|} e^{\mathcal{K}(r_{+})} \bigg( 4 g^2 r_{+} - \frac{\mu}{r_{+}^2} +2 \frac{c}{r_{+}} \bigg) \bigg{|}\,,
\end{equation}
where we used the fact that at the horizon the function $U(r)$ vanishes.

The formula \eqref{temp} is in general cumbersome given the fact that we
first have to find roots of the quartic polynomial. Given the form of the roots, we can rewrite the formula for the temperature in a slightly more convenient way, namely in function of the difference $(r_{+}-r_{-})$. Indeed, the metric \eqref{nonextr_new_sol} can be expressed as
\begin{equation}
U^2(r) =e^{\mathcal{K}} (r-r_{+})(r-r_{-})(x +\frac{y}{r} +\frac{t}{r^2})\,,
\end{equation}
with parameters
\begin{equation}
g^2=x\,, \qquad c=x r_+ r_{-} +t-y(r_{+}+r_{-})\,, \qquad  \mu=   (r_{+}+r_{-})t -r_{+} r_{-} y\,, \qquad Q= r_{+}r_{-}t\,,
\end{equation}
with the constraint $y=x(r_{+}+r_{-})$. Then the temperature is given by
\begin{equation}
T=\frac{1}{4 \pi} |e^{\mathcal{K}(r_{+})} (r_{+}-r_{-})(x +\frac{y}{r_+} +\frac{t}{r_+^2})|\,.
\end{equation}

\end{document}